\documentclass[twocolumn,trackchanges]{aastex7} %,linenumbers

\begin{document}

\title{GrAviPaSt's Lens to the Past:\\Unveiling the Evolution of Filamentary Structures}

\author[orcid=0009-0003-2960-1563,sname='Parsa Ghafour']{Parsa Ghafour}
\affiliation{Department of Astronomy and High Energy Physics, Kharazmi University, Tehran, Iran}
\email[show]{P.Ghafour@outlook.com}
\author[orcid=0000-0003-0126-8554,sname='Saeed Tavasoli']{Saeed Tavasoli}
\affiliation{Department of Astronomy and High Energy Physics, Kharazmi University, Tehran, Iran}
\email[]{tavasoli@ipm.ir}

\begin{abstract}

This paper examines the evolution of cosmic filaments across redshifts 1, 0.5, and 0 using the IllustrisTNG100-1 magneto-hydrodynamical simulation. To achieve this, we introduce GrAviPaSt, a simple, efficient and parameter-free filament identification method that leverages gravitational potential, an A*-like path-finding algorithm, and spanning trees. Applying this method to galaxy distributions at different redshifts allows us to analyze various filament properties, including their length, thickness, mass density contrast, and radial profile. Additionally, we investigate dynamic characteristics such as the mean distance of filament galaxies from the skeleton, their weighted mean velocity, and velocity trends normalized by their positions within the filaments. Our findings reveal the evolution of cosmic filaments from redshift 1 to 0, highlighting key differences across classifications. In particular, we examine the mass density contrast radial profile of filaments connecting two galaxy groups and those linking two galaxy clusters, identifying distinct differences in profile shape between these categories. Furthermore, in the context of weighted mean velocity, we analyze cosmic filaments exhibiting either negative or positive weighted mean velocity, demonstrating their differing evolutionary trends in terms of the mean distance of filament galaxies from the skeleton.

\end{abstract}

%% KEYWORDS !!!!!!

\section{Introduction\label{sec:1}}
The large-scale structure (LSS) of the universe reveals an intricate network known as the cosmic web, composed of clusters (or knots), filaments, walls (or sheets), and vast low-density regions known as cosmic voids (\cite{davis1982survey,bond1996filaments,peebles2020large}). Observations from redshift surveys (e.g., \cite{de1986slice,colless20012df}) and insights from numerical simulations (e.g., \cite{springel2005simulations,vogelsberger2014properties}) contributed significantly to identifying the LSS.

The cosmic web originates from primordial density fluctuations that grew through gravitational instability, causing dark matter to collapse under gravitational forces and forming the filamentary structures observed in the galaxy distribution today (\cite{peebles1967gravitational,zel1970fragmentation,zeldovich1982giant}). Within this structure, matter flows from less dense environments like voids to denser regions such as walls, filaments, and nodes, where it accumulates (\cite{galarraga2022relative}).

Galaxies within the cosmic web are not randomly distributed; their properties, such as mass, activity, morphology and luminosity, are closely correlated with their environment (e.g., \cite{kuutma2017voids,wang2018alignment}). Characterizing the topology, density and dynamical state of the cosmic web is therefore crucial for understanding the evolution of galaxies and refining cosmological models (\cite{santiago2020identification}).

Cosmic filaments are fundamental structures of the LSS, forming a crucial part of the cosmic web and acting as boundaries for voids while bridging galaxy groups and clusters across intergalactic space (\cite{cautun2014evolution,zakharova2023filament,wang2024boundary}). These elongated features span vast distances, channeling matter along their lengths toward galaxy halos and clusters and playing a pivotal role in shaping the universe's LSS (\cite{zeldovich1982giant,tempel2014detecting}).

The filamentary environment significantly influences the evolution of galaxies. Factors such as local environmental density and proximity to the filament's central axis play critical roles in influencing galaxy development (\cite{bonjean2020filament,ganeshaiah2021cosmic}). Studies have demonstrated that galaxies closer to the filament skeleton exhibit higher fractions of passive evolution compared to those situated farther away in the surrounding void regions, highlighting the pivotal role of filaments in regulating galaxy properties and star formation activity (e.g., \cite{singh2020study}).

Filaments not only host galaxies composed of cooler baryonic matter but also serve as conduits for warm-hot intergalactic matter, significantly contributing to the cosmic baryon cycle (\cite{galarraga2022relative}). Also, some large-scale phenomena within filaments, such as cosmic web enhancement and detachment, have been identified and linked to the presence of ionized gas clouds in the circumgalactic environments of certain filament galaxies (\cite{vulcani2019gasp,aragon2019galaxy}). Furthermore, extensive research has investigated the statistical characteristics of filaments, including their lengths, radial density profiles and magnetic field strengths, providing deeper insights into their structure and impact (\cite{malavasi2020characterising,vernstrom2021discovery,yang2022universal,galarraga2022relative}). Studies on hydrodynamical simulations have further illustrated how filaments act as conduits connecting galaxies to clusters, refining our understanding of the complex processes underlying cosmic evolution (e.g., \cite{rost2021threehundred,gouin2022gas}).

Observations of filaments have primarily been derived from galaxy redshift surveys, such as the Sloan Digital Sky Survey (SDSS; \cite{abazajian2009seventh}) and the Dark Energy Spectroscopic Instrument (DESI; \cite{adame2024early}). These observations reveal individual filaments spanning several megaparsecs, forming bridges of matter connecting pairs of galaxy clusters (e.g., \cite{biffi2022erosita,hincks2022high}). Meanwhile, cosmological simulations provide a more detailed understanding of their intricate geometry and their function in channeling matter from voids to galaxy clusters (e.g., \cite{wang2024darkai}). Despite their significant role in galaxy formation and cosmic evolution, the complex and elongated shapes of filaments present challenges for their identification in both observational data and simulations based on dark matter particles (\cite{zakharova2023filament}).

Over the years, a wide range of methods have been developed to identify cosmic filaments by analyzing the spatial distribution of particles in simulations and observational galaxy data. These approaches include geometric and topological evaluations, such as the Bisous model (\cite{tempel2016bisous}), Skeleton (\cite{sousbie20083d}), NEXUS (\cite{cautun2013nexus}) and tessellation-based tools like DisPerSE (\cite{sousbie2011persistent}). Other techniques, including multiscale morphology filters MMF (\cite{aragon2007multiscale,aragon2010multiscale}), segmentation-based models like the Candy model (\cite{stoica2005detection}), and watershed methods such as SpineWeb (\cite{aragon2010multiscale}), offer diverse ways to detect filaments. Each of these methods has unique strengths, yet comparing their effectiveness is complicated by the absence of a universally accepted definition of cosmic filaments and the varying requirements across studies (\cite{alina2022malefisenta,inoue2022classification,zhang2022sconce,aragon2024hierarchical}).

The results produced by filament detection algorithms are highly sensitive to the choice of parameters and identification methods. For example, smoothing scales significantly influence the width and length distributions of filaments identified in density maps, as observed in methods applied to surveys such as SDSS and DEEP2 (\cite{bond2010crawling,choi2010tracing}). Similarly, algorithms like DisPerSE demonstrate how parameters like persistence thresholds can alter filament properties, removing shorter filaments or modifying length distributions (\cite{malavasi2020characterising}). Additionally, the process of determining filament endpoints plays a crucial role, as this directly impacts length measurements (\cite{malavasi2020characterising}). Statistical analyses also reveal that filament characteristics can differ based on the types of tracers used, whether galaxies or dark matter particles.

An effective filament detection algorithm should first and foremost align with human visual perception while providing quantitative and reproducible results rooted in sound numerical theory. Furthermore, it should minimize dependency on adjustable parameters, ensuring reliable outcomes with simplicity. Efficiency is also a critical factor, as the algorithm must achieve high performance without excessive computational demands (\cite{santiago2020identification}).

Finally, studies have highlighted intriguing differences in the radial density profiles of filaments. For example, investigations using the DisPerSE algorithm on simulations like Illustris-TNG show that shorter filaments exhibit distinct density profiles compared to their longer counterparts (\cite{galarraga2022relative}).

In this study, we present a novel yet simple and effective method for detecting cosmic filaments. This method integrates Gravitational potential, an A*-like Path-finding algorithm,  and Spanning trees (GrAviPaSt) to identify filamentary structures within galaxy distributions  as described in Section \ref{sec:3}. We then apply the GrAviPaSt method to galaxy distributions at three distinct redshifts: 0, 0.5, and 1, and present the results in Section \ref{sec:4}. By comparing these findings across different redshifts, we investigate the evolutionary trends of these cosmic filaments over cosmic time. The galaxy distributions used in this study are sourced from the IllustrisTNG100-1 simulation, as detailed in Section \ref{sec:2}. Finally, in Section \ref{sec:5}, we summarize our findings and further discuss the results presented in Section \ref{sec:4}.

\section{Data: Illustris-TNG\label{sec:2}}
The dataset for this study is derived from the gravo-magnetohydrodynamical IllustrisTNG\footnote{https://www.tng-project.org} simulation (\cite{nelson2019illustristng}), which simulates the coupled evolution of dark matter, gas, stars, and black holes throughout cosmic history, spanning redshift z = 127 to z = 0. This simulation employs the Arepo moving-mesh framework (\cite{springel2010pur}) and adopts cosmological parameters informed by the Planck cosmology (\cite{ade2016planck}). These include density parameters of $\Omega_{m,0}$ = 0.3089 for matter, $\Omega_{b,0}$ = 0.0486 for baryons, and $\Omega_{\Lambda,0}$ = 0.6911 for the cosmological constant, along with $n_{s}$ = 0.9667 for the scalar spectral index, $\sigma_8$ = 0.8159 for the amplitude of fluctuations, and $H_{0} = 100 h$ $(km/s/Mpc)$ with $h_{0}$ = 0.6774 for the Hubble constant.

For this analysis, the focus is on the IllustrisTNG100-1 simulation box, which represents the highest-resolution run within the IllustrisTNG100 suite. This box encompasses a cubic volume with a side length of approximately 75 $(Mpc/h)$, featuring a dark matter and baryonic resolution of $7.5\times10^6M_\odot$ and $1.4\times10^6M_\odot$ respectively. This combination of spatial scale and resolution allows for both the reliable identification of extended cosmic filaments and a detailed analysis of their diverse characteristics.

In this study, we selected all simulated galaxies from the illustrisTNG100-1 catalog that are brighter than approximately -16 in the r-band filter, possess stellar masses exceeding $10^8$ $M_\odot$, and exist at three different redshifts. We used the snapshots 50, 67 and 99, respectively corresponding to redshifts 1.0, 0.5 and 0. Using these thresholds, our final samples in these three redshifts contains about $46000$, $45000$ and $41000$ galaxies respectively.
\section{Methodology: GrAviPaSt\label{sec:3}}
In this section, we introduce a novel, simple and yet efficient cosmic filament-finding algorithm called the GrAviPaSt method. This algorithm is specifically designed to operate with a minimal number of parameters and ensuring computational efficiency. Due to its simplicity, the GrAviPaSt method can be applied to large datasets while requiring a reasonable amount of computational resources.

The GrAviPaSt method combines several powerful techniques to achieve its goals. It leverages the minimum spanning tree (MST) algorithm (\cite{karger1995randomized}), integrates the concept of gravitational potential, utilizes an A*-like path-finding algorithm (\cite{hart1968formal}), and takes into account the inherent geometry of the input dataset. This method consists of four main phases, which are applied sequentially to the input dataset. Each phase involves a series of steps.

\subsection{Prim’s MST\label{sec:3.1}}
In the first phase, GrAviPaSt method employs the minimum spanning tree (MST) algorithm to identify pairs of galaxy groups and search for the micro-filament structures (hereafter micro-filaments represent filaments connecting two groups of galaxies) between them.

\begin{figure}
\centering
\includegraphics[width=0.47\textwidth]{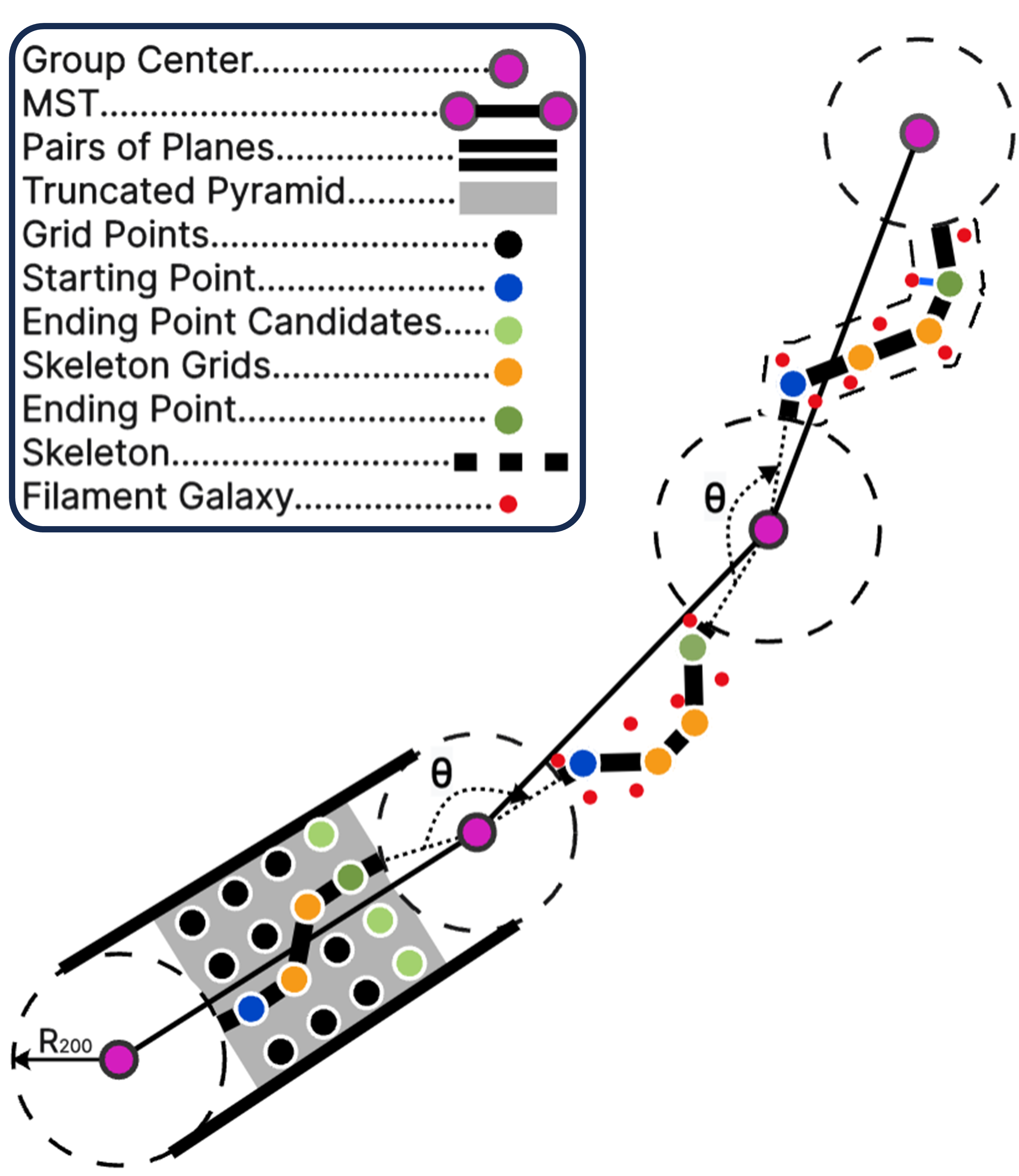}
\caption{
A schematic illustration depicting the sequential phases and steps of the GrAviPaSt method in identifying a macro-filament composed of three micro-filaments. Four galaxy groups are presented, with the method’s phases applied sequentially from bottom to top. The black lines connecting the centers of the galaxy groups illustrate the application of Prim’s MST algorithm, used to identify the galaxy group pairs between which GrAviPaSt searches for micro-filaments. Between the first and second galaxy groups, the schematic depicts the A*-like path-finder algorithm that identifies the path of deepest gravitational potential using grid points. Between the second and third, as well as the third and fourth galaxy groups, filament galaxies and the micro-filament radius are identified. Finally, the angles $\theta$ between the connection points of the micro-filaments attached to each galaxy group are illustrated. These angles are used to chain together micro-filaments in order to construct the macro-filament. \label{fig:1}}
\end{figure}

GrAviPaSt method specifically uses Prim's MST algorithm (\cite{prim1957shortest}) instead of other MST algorithms such as Kruskal's algorithm (e.g., as mentioned in \cite{santiago2020identification}). This choice is due to the optimal implementation of Prim's MST. In GrAviPaSt method, the spatial coordinates of each galaxy group's BGG (or BCG) serve as the node coordinates for the input of the MST algorithm. The Prim's MST algorithm begins by initializing the process with an arbitrary node (vertex) from the input dataset, designating it as part of the growing minimum spanning tree (MST), while assigning an initial cost of infinity to all other nodes except the starting node, which has a cost of zero. Next, it iteratively selects the edge with the smallest weight that connects any node in the MST to a node outside it. For each newly added node, the algorithm updates the costs of its neighboring nodes, ensuring that any smaller edge weight replaces the current cost. This process is repeated until all nodes are included in the MST, thereby creating a structure with minimum-cost connections. Following the implementation of Prim's MST, the pairs of galaxy groups between which the GrAviPaSt method will search for micro-filaments are identified. An example of the implementation of Prim's MST is shown in Figure \ref{fig:1}.

\subsection{A*-like Path-Finder\label{sec:3.2}}
In the second phase, the GrAviPaSt method employs an A*-like path-finding algorithm combined with gravitational potential. This phase consists of multiple sequential steps applied to the input dataset. To implement this phase, the GrAviPaSt method utilizes the output of the MST algorithm from the previous phase.

For each pair of galaxy groups determined by the MST algorithm (Section \ref{sec:3.1}), the GrAviPaSt method first identifies the minimum and maximum coordinates of the galaxies within these two groups. It then crops the dataset box to these coordinates, creating a rectangular cuboid (hereafter referred to as the cropped box), which contains only the galaxies inside this defined space. It is worth noting that the GrAviPaSt method includes all galaxies within the cropped box, not just those belonging to the two galaxy groups.

Next, the GrAviPaSt method centers two spheres on the coordinates of the BGGs (or BCGs) of the two groups, using the virial radius $R_{200}$ of each group as the radius of the spheres. Subsequently, it tangentially places two pairs of planes around these spheres. Each pair of planes is parallel to each other but perpendicular to the other pair of planes. Using these four planes, the GrAviPaSt method constructs a pyramid-shaped structure around the two spheres, with the line connecting the two BGGs (or BCGs) forming the central axis. The algorithm retains only the region that is inside the pyramid, outside the spheres of the galaxy groups, and between the two spheres. The resulting space is a truncated pyramid: its base plane is tangential to the sphere of the larger group, its top plane is tangential to the sphere of the smaller group, and its four trapezoidal sides are defined by the two pairs of planes.

With the truncated pyramid and cropped box in place, the GrAviPaSt method moves to the next step of this phase. In this step, the truncated pyramid is filled with grid points, each separated by a distance of $d$ Mpc from its horizontal and vertical neighbors. The value of d is determined by:
\begin{equation}
\label{eq:1}
d = \sqrt[3]{\frac{N}{V}}
\end{equation}
where $N$ is the total number of galaxies in the input dataset (not the cropped box), and $V$ is the total volume of the input dataset. These grid points are shown between the galaxy group pairs in Figure \ref{fig:1}.

Once the grid points are established, the GrAviPaSt method proceeds to implement the A*-like path-finding algorithm. This involves several steps, including determining the starting point, selecting the ending point candidates, and identifying the path with the minimum gravitational potential to locate the deepest gravitational well between two groups.

To determine the starting point, the GrAviPaSt method identifies grid points adjacent to the top plane of the truncated pyramid as candidates. Thus, the starting point is always selected near the smaller galaxy group. For each starting point candidate, the gravitational potential of the galaxies within the cropped box is calculated using:
\begin{equation}
\label{eq:2}
\Phi = -G\sum_{i}\frac{M_i}{\sqrt{(x-x_i)^2 +(y-y_i)^2+(z-z_i)^2}}
\end{equation}
where the coordinates $x$, $y$, and $z$ correspond to the starting point candidate, and the parameters with the $i$-subscript refer to all galaxies within the cropped box. Among these candidates, the grid point with the lowest (deepest) gravitational potential is chosen as the starting point.

For the ending point candidates, the GrAviPaSt method considers grid points adjacent to the base plane of the truncated pyramid, ensuring that the end of the path is always located near the larger galaxy group. With both the starting point and the ending point candidates established, the A*-like path-finding algorithm can now be applied.

To identify the path with the deepest gravitational potential between two groups, the A*-like path-finding algorithm begins at the starting point. From this point, it considers all neighboring grid points that are closer to the BGG (or BCG) of the larger group than to the starting point. For each of these neighboring grid points, GrAviPaSt calculates the gravitational potential using Equation \ref{eq:2} and selects the grid point with the lowest (deepest) gravitational potential as the next step in the path. This process is repeated iteratively, with each subsequent point added to the path, until the algorithm reaches one of the ending point candidates. Once an ending point candidate is reached, the path-finding process halts.

After determining this minimum path (or deepest gravitational well), the GrAviPaSt method considers this path as the main skeleton of the micro-filament connecting the two galaxy groups. These skeletons are illustrated between galaxy groups in Figure \ref{fig:1}.

\subsection{Micro-Filaments Structure\label{sec:3.3}}
In the third phase, GrAviPaSt method constructs the final structure of the micro-filaments based on the distribution of galaxies surrounding the skeleton (deepest potential path) identified between the two groups of galaxies in the previous phase.

The GrAviPaSt method models micro-filaments as cylindrical structures that bridge the two galaxy groups, with their central axis aligning with the skeleton identified in the previous phase (Section \ref{sec:3.2}). To determine the radius of these micro-filaments, GrAviPaSt examines the galaxies surrounding the skeleton. Specifically, it considers galaxies that lie outside the virial radius of the groups, between the two groups, and within a distance from the skeleton that is less than or equal to the average of the two groups' virial radius $R_{200}$. Among these galaxies, the GrAviPaSt method identifies the galaxy farthest from the skeleton and assigns this distance as the radius of the micro-filament.

Once the final structure of the micro-filaments is determined, the GrAviPaSt method evaluates all the micro-filaments and eliminates those containing only a single galaxy. As a result, all of the final micro-filaments, as well as the macro-filaments (Section \ref{sec:3.4}), contain at least two galaxies within their cylindrical structure. This cylindrical structure is depicted in Figure \ref{fig:1}.

\subsection{Macro-Filaments\label{sec:3.4}}
The final phase involves grouping the micro-filaments like chains to construct the macro-filaments. To achieve this, the GrAviPaSt method follows a three-step process based on the thickness, angle, and gravitational potential of the micro-filaments associated with each node, determining which pairs of micro-filaments connected to a node can be grouped into the same macro-filament.

First, the algorithm calculates the $2\sigma$ (standard deviation) of all micro-filament thicknesses connected to a node, identifying the thickness error for that node. The GrAviPaSt method begins with the thickest micro-filament connected to the node as the base micro-filament and considers other micro-filaments with thicknesses within the $\pm$ error range of the base micro-filament as pair candidates.

Second, among these pair candidates, the algorithm calculates the angle between each candidate and the base micro-filament. To do so, the GrAviPaSt method uses the coordinates of the starting or ending points of the pair candidates corresponding to the node. By connecting these points to the node, the algorithm calculates the angle between the candidates and the base micro-filament. It always considers the smaller angle between the pair candidates and the base micro-filament. The GrAviPaSt method then retains only those pair candidates with an angle greater than 90 degrees, removing the others from consideration.

In the final step, if multiple pair candidates meet the thickness and angle criteria, the GrAviPaSt method uses gravitational potential to determine the best pair. It calculates the difference between the gravitational potential of the points of the pair candidates connected to the node and that of the base micro-filament. The algorithm selects the micro-filament with the lowest potential difference as the best pair. If no suitable pair is found for the base micro-filament, the algorithm moves to next thickest micro-filament connected to the node and considers it as the base micro-filament, repeating the process until a pair is identified or all micro-filaments associated with the node are examined without finding a pair. After determining the best pair for each node, the algorithm groups paired micro-filaments together like a chain, forming the final structure of the macro-filaments. These macro-filaments can consist of one or more micro-filaments.

\section{Results\label{sec:4}}
This section presents the findings derived from the application of the GrAviPaSt method to the three datasets described in Section \ref{sec:2}. By identifying both micro- and macro-filaments using this method, we analyze their various characteristics across different redshifts: 1, 0.5, and 0, covering approximately the last 8 Gyrs.

For the Prim’s MST phase of the GrAviPaSt method, the coordinates of BGGs (or BCGs) are chosen from galaxy groups containing at least 5 members. These selected coordinates serve as the nodes for the MST algorithm, with the micro-filaments identified in this study forming bridges between them.

\subsection{Micro-Filaments\label{sec:4.1}}
In this subsection, we examine the characteristics of the micro-filaments. First, we analyze the length and thickness distributions of these filaments, providing insights into their structural properties. Subsequently, we investigate the respected velocity of filament galaxies along the micro-filament skeleton, revealing their dynamic behavior. Finally, we explore the radial profiles of the micro-filaments to identify patterns in their density and structural composition.
\subsubsection{Length Evolution\label{sec:4.1.1}}
One of the primary characteristics analyzed is the length distribution of micro-filaments across different redshifts. The length of these micro-filaments is determined by summing the distances between neighboring grid points that form the micro-filament skeleton (Section \ref{sec:3.2}). 

\begin{figure}
\centering
\includegraphics[width=0.42\textwidth]{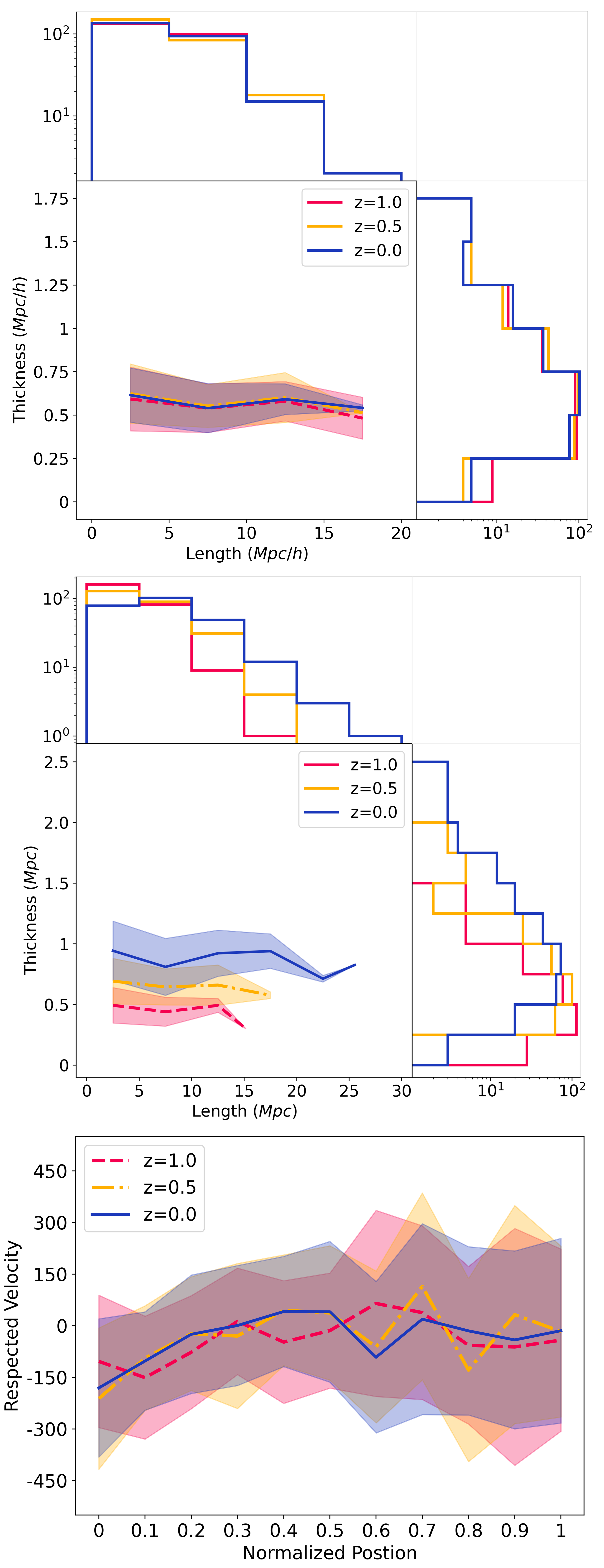}
\caption{
The median trend of micro-filament thickness, along with its associated error, is presented for their length distribution in co-moving (\textit{Top}) and proper (\textit{Middle}) coordinates. At the \textit{Bottom}, the median trend of the velocity of galaxies along the micro-filaments' skeleton is shown. Note that, according to the implementation of the GrAviPaSt method, the skeleton of micro-filaments originates at the smaller group and terminates at the larger group.
\label{fig:2}}
\end{figure}

The distribution of micro-filament lengths is illustrated in Figure \ref{fig:2}, with the top panel showing the co-moving lengths, and middle panel the proper lengths. As shown in this figure, the co-moving lengths of micro-filaments remain within a similar range across different redshifts, exhibiting slow evolution. By factoring out the Hubble flow, the specific influences of gravity and DE on filament length evolution become more apparent (\cite{galarraga2024evolution}).

In contrast, the distribution of proper filament lengths varies significantly across different redshifts. As illustrated in the middle panel of Figure \ref{fig:2}, the range of micro-filament lengths expands with decreasing redshift. This global growth, evident in proper coordinates, is a direct consequence of the Universe’s expansion. Driven by the Hubble flow, the cosmic web undergoes continuous stretching, leading to an overall increase in micro-filament lengths.

\begin{deluxetable*}{ccccc}  
\tablewidth{0pt}
\tablecaption{P-values from two-sided Kolmogorov–Smirnov tests comparing the distributions of micro-filament characteristics between pairs of redshifts. \label{tab:a}}
\tablehead{
	\multicolumn{1}{c}{Redshift pairs} & \multicolumn{1}{c}{} & \multicolumn{1}{c}{0 and 0.5} & \multicolumn{1}{c}{0 and 1} & \multicolumn{1}{c}{0.5 and 1}
}  
\startdata
Length & Co-moving & 0.32 & 0.02 & 0.16 \\
$ $ & Proper & $1.46\times10^{-7}$ & $5.13\times10^{-29}$ & $3.15\times10^{-11}$ \\
Thickness & Co-moving & $9.03\times10^{-3}$ & 0.41 & 0.03 \\
$ $ & Proper & $1.18\times10^{-4}$ & $3.44\times10^{-36}$ & $2.19\times10^{-25}$ \\
\enddata
\end{deluxetable*}

The statistical analysis, performed using two-sided Kolmogorov–Smirnov (K–S) tests, further supports these findings. The resulting p-values, summarized in Table \ref{tab:a}, correspond to comparisons of the co-moving length distributions of micro-filaments at redshifts 0 and 0.5, 0 and 1, and 0.5 and 1. These results indicate no statistically significant differences between the distributions at redshifts 0 and 0.5, as well as 0.5 and 1, but show a significant difference between redshifts 0 and 1. In contrast, comparisons of the proper length distributions, also yield p-values listed in Table \ref{tab:a}, which indicate statistically significant differences among these distributions. Additionally, Figure \ref{fig:2} illustrates that the micro-filament population, in both proper and co-moving coordinates, is consistently dominated by shorter filaments across all redshifts.

\subsubsection{Thickness Evolution\label{sec:4.1.2}}
The thickness distribution of micro-filaments across different redshifts, in both co-moving and proper coordinates, represents another key characteristic examined in this study. As illustrated in the panels of Figure \ref{fig:2}, the range of thickness distributions in co-moving coordinates remains largely consistent across redshifts, similar to the length distributions. However, in proper coordinates, as redshift decreases, the range of thickness distributions expands. This thickening of micro-filaments over cosmic time, akin to the trend observed in proper length distributions, is driven by the Hubble flow and the Universe’s continuous expansion.

Further statistical analysis using K–S tests reinforces these findings. The resulting p-values, summarized in Table \ref{tab:a}, correspond to comparisons of the co-moving thickness distributions at redshifts 0 and 0.5, 0 and 1, and 0.5 and 1. These values indicate statistically significant differences between the distributions at redshifts 0 and 0.5, as well as 0.5 and 1, but no significant difference between redshifts 0 and 1, in contrast to the observations for micro-filament length distributions (\ref{sec:4.1.1}). Comparisons in proper coordinates also yield p-values listed in Table \ref{tab:a}, demonstrating statistically differences among these distributions.
\subsubsection{Filament Galaxies Respected Velocity\label{sec:4.1.3}}
As another characteristic of micro-filaments, we examine the velocities of filament galaxies relative to the micro-filament skeleton, accounting for their normalized positions along the micro-filament. To achieve this, we identify the nearest grid point on the micro-filament skeleton for each galaxy. The length of the micro-filament from its starting point to this grid point is calculated (Section \ref{sec:4.1.1}) and then normalized by the total length of the micro-filament to determine the galaxy's normalized position.

It is important to note that the micro-filament skeletons, as identified by the GrAviPaSt method, begin at the smaller galaxy group and extend to the larger one (Section \ref{sec:3.2}). Smaller normalized positions correspond to galaxies near the smaller group, while larger normalized positions indicate galaxies closer to the larger group. In the bottom panel of Figure \ref{fig:2}, negative velocities represent galaxies moving toward the micro-filament skeleton, while positive velocities indicate galaxies moving away from it. As shown in this figure, the median of the respected velocity of filament galaxies remains close to zero across different redshifts. The trend is nearly identical at redshifts 1 and 0.5, but a subtle shift toward more negative velocities is observed from redshift 0.5 to 0.

\subsubsection{Radial Profiles\label{sec:4.1.4}}
The final attribute examined for micro-filaments is their radial profile. Given that the GrAviPaSt method identifies micro-filaments based on the gravitational potential of galaxies (Section \ref{sec:3}), we explored the variations in mass density contrast along their radii. To derive this profile, we first measured the perpendicular distance of each filament galaxy from the micro-filament skeleton and then normalized it to the micro-filament's radius. With these normalized distances, we then combined all filament galaxies, stacking them to create a comprehensive micro-filament that integrates galaxies across different radial positions. The mass density contrast was then determined at various normalized radial distances from the skeleton by:
\begin{equation}
\label{eq:3}
\delta_{m}(r) = \frac{\rho_{m}(r)-\rho_{m}^{b}}{\rho_{m}^{b}}
\end{equation}
where $\rho_{m}(r)$ represents the mass density at a normalized radius $r$, $\delta_{m}(r)$ denotes the corresponding mass density contrast, and $\rho_{m}^{b}$ refers to the background mass density.

To model and compare these radial profiles across different redshifts, we applied the generic-$\beta$ model from the Plummer profile (\cite{cavaliere1976x,arnaud2009beta,ettori2013mass,galarraga2022relative}). This profile is expressed as:
\begin{equation}
\label{eq:4}
1+\delta_{m}(r) = \frac{1+\delta_{m}^{1}}{(1+r^{\alpha})^{\beta}}
\end{equation}
where $\delta_{m}(r)$ is from Equation \ref{eq:3}, and $\delta_{m}^{1}$ represents the mass density contrast at $r = 1$. Given that our analysis focuses on logarithmic scales, we reformulated this profile as:
\begin{equation}
\label{eq:5}
log_{10} (\delta_{m}(r)) = log_{10}\left(\frac{1+\delta_{m}^1}{(1+r^{\alpha})^{\beta}} - 1\right) + \gamma
\end{equation}
where $\alpha$, $\beta$, and $\gamma$ serve as fitting parameters.

\begin{figure}
\centering
\includegraphics[width=0.43\textwidth]{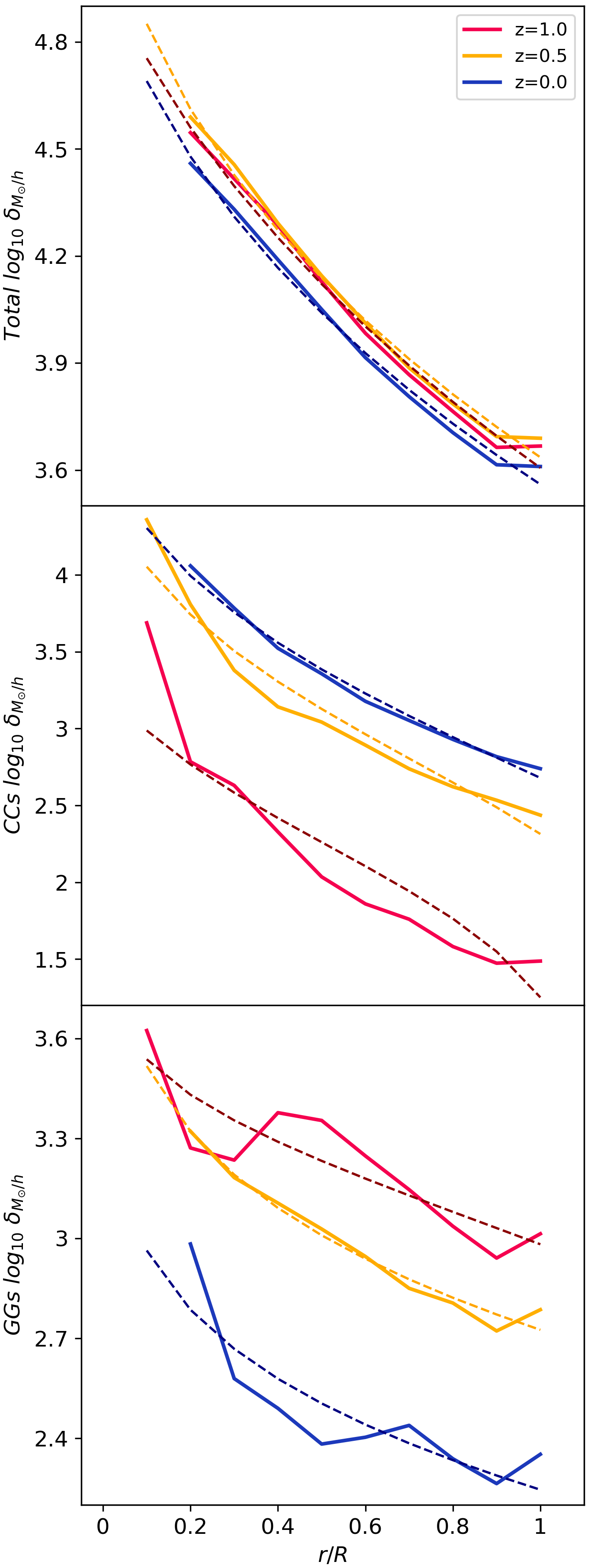}
\caption{
The radial profiles of mass density contrast are presented for total (\textit{top}), CC (\textit{middle}) and GG micro-filaments (\textit{bottom}), each displayed across three redshifts using different colors. The corresponding fit curves, based on the Plummer profile, are represented by dashed lines.
\label{fig:3}}
\end{figure}

We selected this profile due to its effective representation of the total mass density contrast of micro-filaments. The radial mass density contrast profile of micro-filaments across three different redshifts is illustrated in the top panel of Figure \ref{fig:3}, with the corresponding fit-curves shown as dashed lines. The fitted parameters and Mean Squared Errors (MSEs) are summarized in Table \ref{tab:1}. As depicted in this figure, the mass density contrast declines as the distance from the micro-filament skeleton increases. Additionally, as redshift decreases, micro-filaments exhibit a gradual reduction in mass density contrast, reflecting a slow evolutionary trend.

Beyond the total population of micro-filaments, we also examined two specific types: those connecting galaxy clusters (CCs) and those linking groups of galaxies (GGs). CC micro-filaments were identified as structures bridging galaxy groups with at least 50 members, while GG micro-filaments connected groups containing 5 to 10 members. The middle and bottom panels of Figure \ref{fig:3} illustrate the profiles and corresponding fit-curves for CC and GG micro-filaments, modeled using the Plummer profile. Their fitted parameters and MSEs are detailed in Table \ref{tab:1}.

As shown in Figure \ref{fig:3}, the radial mass density profiles of CC and GG micro-filaments across all three redshifts differ significantly from those of total filaments, indicating distinct evolutionary trajectories for these micro-filament types from redshift 1 to 0.

\begin{deluxetable*}{ccccccccccccc}  
\tablewidth{0pt}  
\tablecaption{The fitted parameters of the Plummer profile, as defined in Equation \ref{eq:5}, along with the Mean Squared Errors (MSEs) of the corresponding fit curves. \label{tab:1}}  
\tablehead{  
	\multicolumn{1}{c}{Redshift} & \multicolumn{4}{c}{z=0} & \multicolumn{4}{c}{z=0.5} & \multicolumn{4}{c}{z=1}\\  
	\colhead{Parameter} & \colhead{$\alpha$} & \colhead{$\beta$} & \colhead{$\gamma$} & \colhead{$MSE$} & \colhead{$\alpha$} & \colhead{$\beta$} & \colhead{$\gamma$} & \colhead{$MSE$} & \colhead{$\alpha$} & \colhead{$\beta$} & \colhead{$\gamma$} & \colhead{$MSE$}
}  
\startdata
Total & 0.73 & 4.95 & 1.44 & 0.0006 & 0.68 & 5.55 & 1.62 & 0.0007 & 0.84 & 4.71 & 1.36 & 0.0008 \\  
CCs & 0.63 & 7.15 & 2.22 & 0.0016 & 0.61 & 7.08 & 2.31 & 0.0173 & 0.71 & 4.68 & 1.88 & 0.0739 \\  
GGs & 0.29 & 5.31 & 1.57 & 0.0096 & 0.34 & 5.51 & 1.63 & 0.0008 & 0.00015 & 10.01 & 6.13 & 0.0085 \\  
\enddata
\end{deluxetable*} 

\subsection{Macro-Filaments\label{sec:4.2}}
This subsection examines the properties of macro-filaments identified by the GrAviPaSt method. As outlined in Section \ref{sec:3.4}, this method links micro-filaments using a three-step procedure that considers their thickness, angular alignment, and gravitational potential. In this analysis, we explore the characteristics of these macro-filaments and their evolution across different redshifts. Specifically, we investigate their length distribution, mass density contrast, and the average distance between filament galaxies and the skeleton of the macro-filaments. These parameters are expressed in proper coordinates rather than co-moving coordinates to incorporate the effects of Hubble flow and the expansion of the Universe.

\subsubsection{Effective Length Evolution\label{sec:4.2.1}}
The effective length (hereafter referred to as length) of macro-filaments is calculated as the cumulative sum of the lengths of all micro-filaments that constitute the macro-filament, while not considering the diameters of the galaxy groups within it.

As observed, the range of length distributions differs across redshifts, with macro-filaments gradually increasing in length as redshift decreases. In all three redshifts, shorter macro-filaments make up the majority of the population; however, their dominance diminishes as redshift decreases. Kernel Density Estimation (KDE) curves are overlaid on the length histograms, showing peaks at 5.59, 4.54, and 2.87 for redshifts 0, 0.5, and 1, respectively. These results indicate evolution in the length distributions over cosmic time. Additionally, a two-sided K-S test conducted on the length distributions of macro-filaments between redshifts 0 and 0.5, 0 and 1, and 0.5 and 1. The resulting p-values, summarized in Table \ref{tab:b}, indicate statistically significant differences among these distributions.

\begin{figure*}
\centering
\includegraphics[width=1\textwidth]{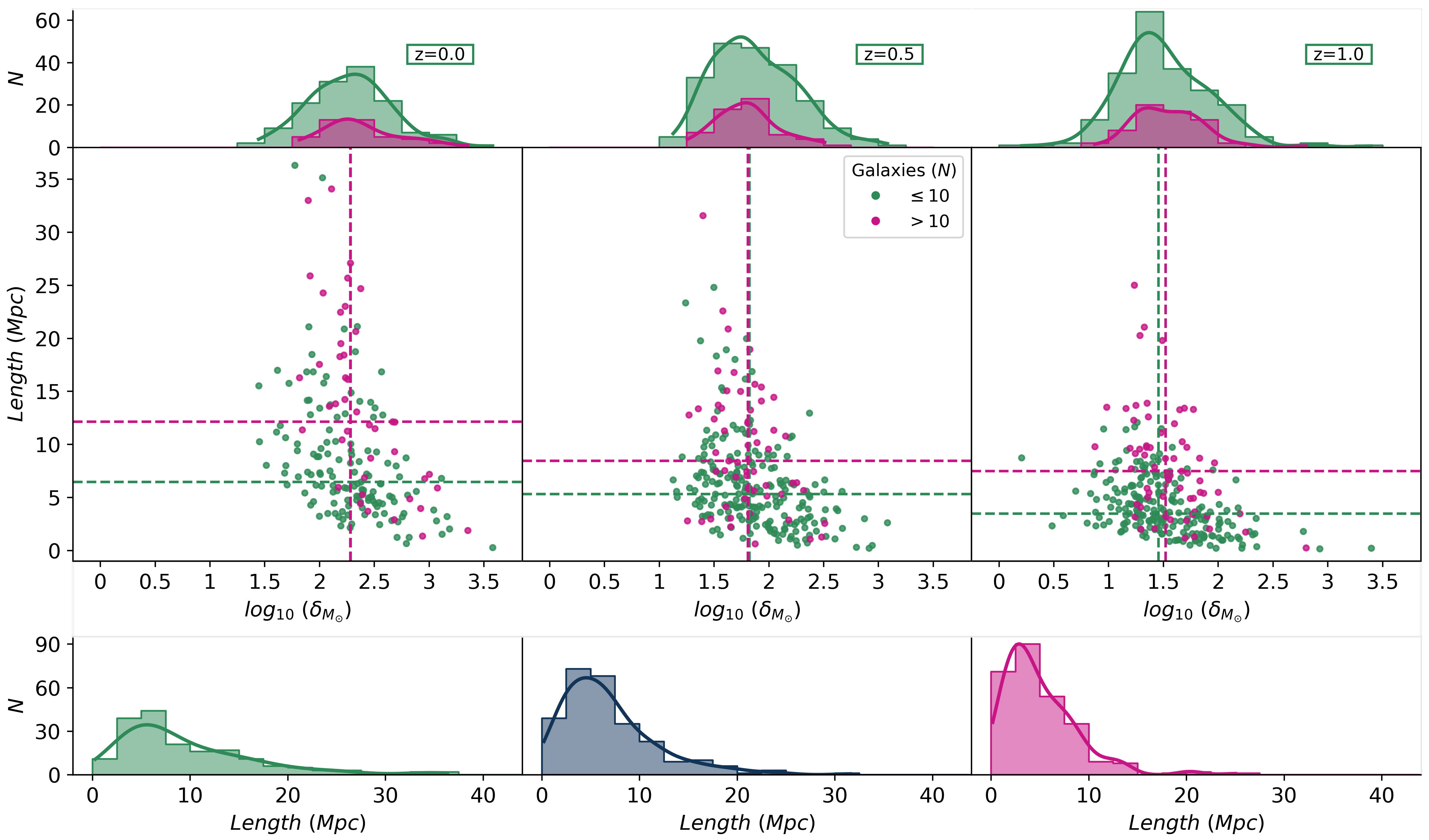}
\caption{
The distribution of macro-filaments length in proper coordinates and mass density contrast across three redshifts is shown. Scatter plots and the mass density contrast histograms use distinct colors to represent the two macro-filament populations, categorized by the number of filament galaxies they host. Horizontal lines mark the median length within each population's distribution, while vertical lines represent the median mass density contrast.
\label{fig:4}}
\end{figure*}

\begin{deluxetable*}{ccccc}  
\tablewidth{0pt}
\tablecaption{P-values from two-sided Kolmogorov–Smirnov tests comparing the distributions of macro-filament characteristics between redshift pairs. Here, $N$ and $\bar{V}$ are the population count and the weighted mean velocity of filament galaxies, respectively. \label{tab:b}}
\tablehead{
	\multicolumn{1}{c}{Redshift pairs} & \multicolumn{1}{c}{} & \multicolumn{1}{c}{0 and 0.5} & \multicolumn{1}{c}{0 and 1} & \multicolumn{1}{c}{0.5 and 1}
}  
\startdata
Effective length & $ $ & $2.21\times10^{-3}$ & $1.47\times10^{-11}$ & $1.83\times10^{-5}$ \\
Mass density contrast & $N \leq 10$ & $1.51\times10^{-17}$ & $4.02\times10^{-40}$ & $3.29\times10^{-13}$ \\
& $N > 10$ & $6.93\times10^{-11}$ & $2.11\times10^{-19}$ & $2.86\times10^{-7}$ \\
Mean distance to skeleton & $\bar{V} \leq 0$ & $2.11\times10^{-3}$ & $8.84\times10^{-19}$ & $4.39\times10^{-14}$ \\
& $\bar{V} > 0$ & $5.52\times10^{-5}$ & $1.13\times10^{-19}$ & $2.71\times10^{-16}$ \\
\enddata
\end{deluxetable*}

The population of macro-filaments can be categorized in multiple ways. Based on length, the population is dominated by shorter macro-filaments, similar to the micro-filaments (see Section \ref{sec:4.1.1}). Another categorization considers the number of filament galaxies hosted by macro-filaments, dividing them into two groups accordingly. These groups are represented by different colors in Figure \ref{fig:4}. Macro-filaments with a low galaxy count ($\leq10$) undergo slow evolutionary changes over cosmic time and consistently maintain shorter lengths across all redshifts. As depicted in Figure \ref{fig:4}, both groups exhibit a gradual increase in proper length, though the gap between their median values widens as redshift decreases.

\subsubsection{Mass Density Contrast Evolution\label{sec:4.2.2}}
Another key characteristic of macro-filaments is their mass density contrast. This metric is preferred over the traditional number density (e.g., investigated in \cite{zhang2024statistical,galarraga2024evolution}) due to the implementation of the GrAviPaSt method, which relies on the gravitational potential of the galaxy distribution to identify the structures of both micro- and macro-filaments. To investigate this property, the mass density of each macro-filament is first computed by summing the total stellar mass of the galaxies hosted within the macro-filament and dividing it by the total volume of the macro-filament. The mass density contrast is then determined using the density contrast formula (Equation \ref{eq:3}).

As shown in Figure \ref{fig:4}, the mass density contrast distribution is more concentrated at redshift 1 and becomes increasingly dispersed as redshift decreases. The median value of each group consistently rises with decreasing redshift. Furthermore, the difference between the medians of the two groups, which is evident at redshift 1, disappears at redshift 0.5, and the medians remain close at redshifts 0.5 and 0. This trend suggests that the poor population group undergoes a greater increase in mass density contrast compared to the rich population group.

A statistical analysis using two-sided K–S tests was conducted on the mass density contrast distributions of macro-filaments with a lower galaxy population ($\leq10$). The resulting p-values, presented in Table \ref{tab:b}, correspond to comparisons between redshifts 0 and 0.5, 0 and 1, and 0.5 and 1, and indicate statistically significant difference between distributions at different redshifts. For macro-filaments containing a larger galaxy population ($>10$), the same K–S tests yield p-values also listed in Table \ref{tab:b}, which similarly confirm significant differences in the mass density contrast distributions across all three redshifts.

\begin{deluxetable}{cccc}
\tablewidth{0pt}
\tablecaption{Total luminosity ($log_{10}$ $L_\odot$), mass ($log_{10}$ $M_\odot$) and volume of the macro-filaments in both co-moving ($(Mpc/h)^3$) and proper ($Mpc^3$) coordinates through different redshifts. \label{tab:2}}
\tablehead{
	\colhead{Redshift} & \colhead{z=0} & \colhead{z=0.5} & \colhead{z=1}
}
\startdata
Total mass & $1.51\times10^{4}$ & $1.96\times10^{4}$ & $1.88\times10^{4}$ \\
Total luminosity & $1.53\times10^{4}$ & $2.05\times10^{4}$ & $2.00\times10^{4}$ \\
Total proper volume & 950.29 & 658.44 & 227.63 \\
Total co-moving volume & 295.39 & 467.16 & 397.89 \\
\enddata
\end{deluxetable}

\subsubsection{Filament Galaxies Mean Radial Distance\label{sec:4.2.3}}
\begin{figure*}
\centering
\includegraphics[width=1\textwidth]{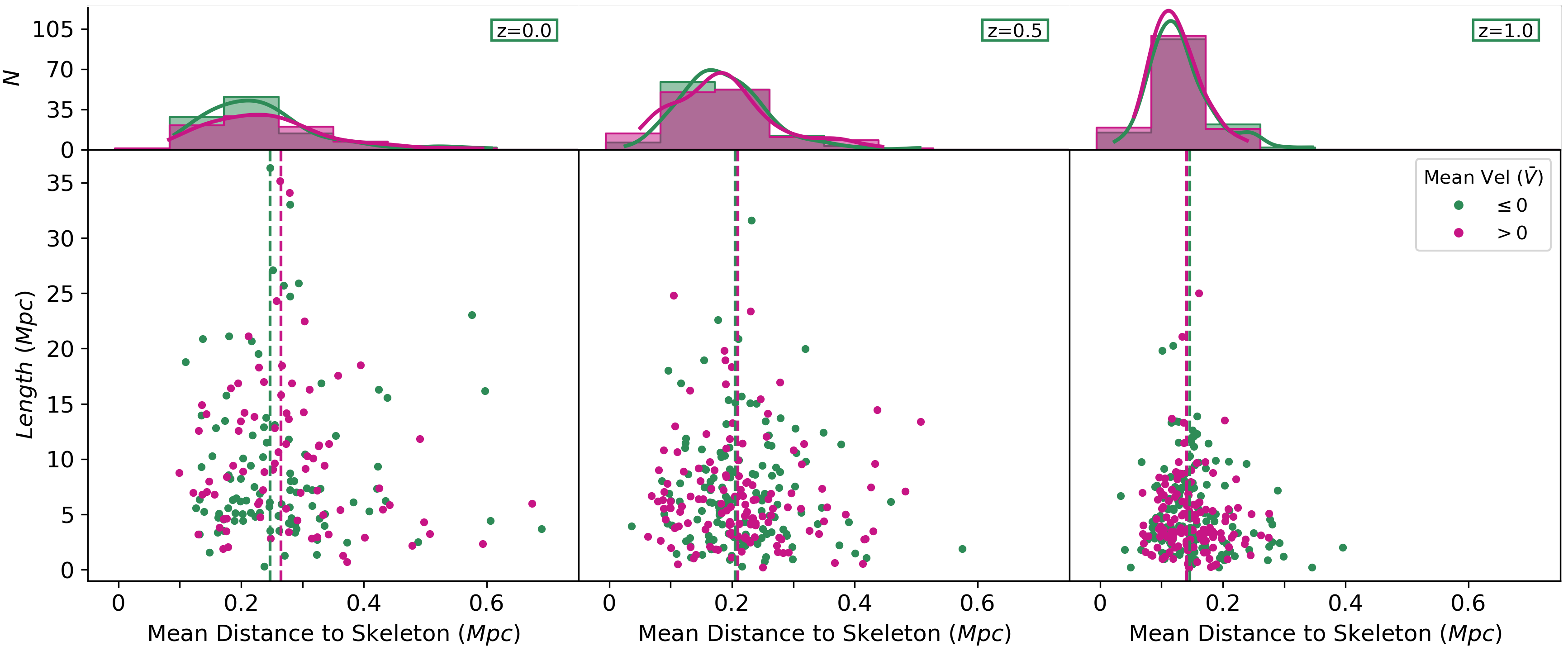}
\caption{
The distribution of the mean distance to the skeleton of macro-filaments across different redshifts is shown with respect to the lengths of the macro-filaments. Macro-filaments with negative and positive weighted mean velocities of their hosted galaxies are depicted using distinct colors.
\label{fig:5}}
\end{figure*}

The next characteristic of macro-filaments examined is the distribution of the mean distance and weighted mean velocity of filament galaxies relative to the skeleton identified by the GrAviPaSt method (Section \ref{sec:3.2}). To compute these parameters, the perpendicular distance between each filament galaxy and the associated skeleton is measured and defined as the galaxy's radial distance. The velocity of the galaxies is calculated following the details outlined in Section \ref{sec:4.1.3}, with the modification of using the stellar mass of galaxies as weights for the weighted mean. For each macro-filament, the mean values of these parameters are then determined. These results are illustrated in Figure \ref{fig:5}, plotted against the lengths of the macro-filaments.

As shown in Figure \ref{fig:5}, the mean distance distribution of filament galaxies expands as redshift decreases. At redshift 1, the distribution is more concentrated, but as redshift decreases, it becomes more dispersed, covering a broader range. To further analyze these dynamics, macro-filaments are classified based on the weighted mean velocity of their hosted galaxies. Two categories are defined: macro-filaments with a negative weighted mean velocity, indicating a predominance of matter infall toward the filament skeleton, and macro-filaments with a positive weighted mean velocity, signifying a predominance of outflow motion away from the skeleton. As depicted in Figure \ref{fig:5}, the median values of both groups increase with decreasing redshift, and the gap between their medians becomes more distinct over cosmic time. The fraction of macro-filaments exhibiting a negative weighted mean velocity is 0.56, 0.49, and 0.5 for redshifts 0, 0.5, and 1, respectively. The stability of this fraction across redshifts suggests slow dynamical evolution of macro-filaments over cosmic time. Additionally, the mean of weighted mean velocities for the negative group is -169.27 $km/s$, -150.09 $km/s$, and -151.06 $km/s$ at redshifts 0, 0.5, and 1, respectively, whereas for the positive group, it is 181.51 $km/s$, 172.73 $km/s$, and 164.37 $km/s$ across these redshifts. The stability of mean values within both groups across different redshifts suggests a slow dynamical evolution of macro-filaments.

Two-sided K–S tests were applied to analyze the distributions of mean distance to the skeleton for both the negative and positive groups, across redshift pairs 0 and 0.5, 0 and 1, and 0.5 and 1. The resulting p-values, as outlined in Table \ref{tab:b}, indicate statistically significant differences in the distribution of both groups across different redshifts.

\subsubsection{Total Luminosity, Mass and Volume Evolution\label{sec:4.2.4}}
Finally we investigate the evolution of the total luminosity and mass of filament galaxies hosted by macro-filaments, as well as the total volume of the macro-filaments in both the proper and co-moving coordinates. The specific details of these parameters are presented in Table \ref{tab:2}.

Based on the data in Table \ref{tab:2}, the total volume of macro-filaments steadily increases from redshift 1 to 0 in proper coordinates. In co-moving coordinates, the total volume increases from redshift 1 to 0.5 and then decreases from redshift 0.5 to 0. This growth in proper coordinates corresponds to the thickening of micro-filaments across these redshifts (see Section \ref{sec:4.1.2}). Additionally, the total mass and luminosity of filament galaxies follow a similar trend to the total co-moving volume, increasing between redshifts 1 and 0.5 and then decreasing between redshifts 0.5 and 0.

\section{Summary and Discussion\label{sec:5}}
In this study we presented GrAviPaSt, a fully deterministic filament-finding algorithm with no free parameters. Its adaptive nature allows it to respond dynamically to the structure and local conditions of the input dataset, producing micro-filaments that reflect the underlying gravitational environment without external tuning. Each phase of the algorithm is directly informed by the data. In the first phase (Section \ref{sec:3.1}), Prim’s MST operates solely based on the spatial configuration of the BGG (or BCG) centers. In the second phase (Section \ref{sec:3.2}), the A*-like path-finding algorithm depends both on the input dataset, which determines the grid spacing $d$, and on the local geometry of each galaxy group (or cluster) pair, which specifies the coordinate ranges of the corresponding cropped box. Since the cropped box dimensions are determined by the spatial extent of group or cluster members, the gravitational potential computed within each cropped box is unique, and consequently, the traced potential-minimizing path varies from one micro-filament to another. This local responsiveness continues in the third phase (Section \ref{sec:3.3}), where the search for filament galaxies is carried out around the skeleton of each micro-filament, with the search radius determined by using the $R_{200}$ values of the associated group (or cluster) pair. The final radial extent of each micro-filament is then determined by the distribution of galaxies along its skeleton, ensuring environment-specific adaptation. Lastly (Section \ref{sec:3.4}), the grouping of micro-filaments into macro-filaments proceeds by evaluating the positional, geometric, and gravitational potential characteristics of all micro-filaments connected to each node (group or cluster). This enables the algorithm to adapt to the local environment and robustly identify coherent pairs of micro-filaments to be linked together. Together, these fully data-driven procedures demonstrate the robustness of GrAviPaSt’s results and emphasize the method’s capacity to adapt naturally to diverse cosmic environments.

The analyses presented in this study revealed that both micro- and macro-filaments exhibit consistent evolution across redshifts 1, 0.5, and 0. The observed distributions of filaments length (Section \ref{sec:4.1.1} and Section \ref{sec:4.2.1}), thickness (Section \ref{sec:4.1.2}), mass density contrast (Section \ref{sec:4.2.2}) and radial profile (Section \ref{sec:4.1.4}) are in general agreement with previous studies (e.g., \cite{galarraga2024evolution,galarraga2020populations}) that employed alternative filament identification algorithms, such as COWS (\cite{pfeifer2022cows}), NEXUS (\cite{cautun2013nexus}), and DisPerSE (\cite{sousbie2011persistent}). These findings reinforce the reliability of GrAviPaSt method in identifying and characterizing cosmic filaments.

In addition, the various characteristics of the filaments provide valuable insights into their dynamics. For instance, the velocity patterns of filament galaxies along the length of micro-filaments. Moreover, the mean weighted velocity of these galaxies highlights the overall inflow and outflow of matter relative to the filament skeleton. Additionally, the relatively close mean distance of galaxies to the skeleton for both categories of macro-filaments across different redshifts emphasizes the structural consistency of these filamentary structures (Section \ref{sec:4.2.3}).

These analyses aimed to assess the effectiveness of the GrAviPaSt method in identifying filaments and to uncover new insights into the evolutionary trends of cosmic filaments from redshift 1 to 0. Nevertheless, it remains challenging to fully evaluate the success of any filament identification approach due to the absence of a detailed consensus on the definition of cosmic filaments, as different studies have adopted diverse methods and criteria for defining these structures. Also, these analyses have helped illustrate various aspects of cosmic filaments and their dynamics. However, further investigation of these characteristics remains necessary in future studies.

The insights presented in this study also offer valuable metrics for evaluating and comparing different cosmological simulations and observations. These include simulations based on the $\Lambda$CDM cosmology employing semi-analytical galaxy formation models (e.g., \cite{henriques2020galaxies, ayromlou2021galaxy}), gravo-magnetohydrodynamical frameworks (e.g., \cite{nelson2019illustristng}), as well as observational data from redshift surveys (e.g., \cite{almeida2023eighteenth, adame2024early}). Parameters such as the lengths of micro-filaments and the effective lengths of macro-filaments, the evolution of macro-filament mass density contrast, radial profiles of density contrast along micro-filament radii, galaxy population counts and the mean distances of galaxies to the filament skeleton within macro-filaments can serve as testable metrics for assessing how accurately simulations reproduce the observed Universe. In particular, tracing the redshift evolution of these parameters and subsequently comparing them across simulations and observational datasets, has the potential to yield valuable insights into possible discrepancies and to refine our understanding of large-scale structure evolution. Furthermore, since the GrAviPaSt method relies on the gravitational potential of the dataset, it may also be suitable for evaluating alternative cosmological models through simulations of dark matter particle distributions and their evolution across redshift (e.g., \cite{lague2024cosmological}).

An alternative and widely used approach to the cosmic web classification, based on the gravitational potential of the dataset, involves analyzing the deformation tensor (i.e., the Hessian of the gravitational potential) and offers a complementary framework to GrAviPaSt. For example, the T-web method (e.g., \cite{forero2009dynamical,ayccoberry2024theoretical}) utilizes the eigenvalues of the deformation tensor to classify the large-scale environment into voids, sheets, filaments, and knots. This technique relies on the local curvature of the potential field to infer structural morphology and identifies filament cells as those with exactly two eigenvalues above a fixed threshold.

While GrAviPaSt also relies on the gravitational potential field, it differs conceptually and operationally from Hessian-based methods. Rather than classifying the cosmic web based on local curvature or eigenvalue thresholds, GrAviPaSt constructs micro-filament skeletons by identifying the path of deepest gravitational potential between galaxy groups. This approach is more trajectory-based and emphasizes the connectivity and continuity of filamentary structures. In this sense, GrAviPaSt complements Hessian-based methods by offering a path-centric, potential-guided reconstruction of filaments, which may be particularly useful for studying their dynamical evolution.

In conclusion, the findings demonstrate that the evolution and characteristics of micro- and macro-filaments identified using the GrAviPaSt method are in general consistent with existing literature while also highlighting areas for future exploration to enhance the study of cosmic structures.

\begin{acknowledgments}
We extend our sincere appreciation to the anonymous referee for their thoughtful evaluation and constructive critique, which have significantly enhanced the clarity, depth, and scholarly rigor of this work.
\end{acknowledgments}

\bibliography{GrAviPaSt}{}
\bibliographystyle{aasjournalv7}

\end{document}